\documentclass[english]{article}
\usepackage[T1]{fontenc}
\usepackage[latin9]{inputenc}
\usepackage[letterpaper]{geometry}
\usepackage{amsmath}
\usepackage{amssymb}
\usepackage{esint}
\usepackage{setspace}
\makeatletter
\usepackage{ae}
\usepackage{aecompl}

\usepackage{babel}
\usepackage[sf,md]{titlesec}

\usepackage[footnotesize,sf,hang]{caption}

\usepackage[multiple]{footmisc}

\usepackage{color} % in case someone forgot
\definecolor{red}{rgb}{0.3,0,0} % used for eq and section refs
\definecolor{green}{rgb}{0,0.3,0} % used for citations
\definecolor{magenta}{rgb}{0,0,0.3} % magenta used for URL links

\usepackage[colorlinks,bookmarks=false]{hyperref} % ,linktocpage=true
\usepackage{color} % in case someone forgot
\usepackage[nosort]{cite} 

\DeclareMathOperator{\e}{\mathbf{e}}
\DeclareMathOperator{\ot}{\otimes}

\DeclareMathOperator{\tr}{tr}

\makeatother

\usepackage{babel}

\begin{document}
\begin{flushright}
arXiv:0912.3468\\
%BROWN-HET 15??\\
\bigskip\bigskip
\par\end{flushright}

\setstretch{2.3}
\begin{center}
\textsf{\huge Poisson Structures of Calogero-Moser and Ruijsenaars-Schneider Models}
\par\end{center}{\Large \par}

\setstretch{1.15}
\begin{center}
\bigskip \textbf{\large In\^es Aniceto,$^{l}$ Jean Avan$^{c}$} and \textbf{\large Antal Jevicki$^{p}$}\\ \emph{ \smallskip }\\
\emph{$^{l}$ CAMGSD, Departamento de Matem\'atica, Instituto Superior T\'ecnico,\\ Av. Rovisco Pais, 1049-001 Lisboa, Portugal}\\ 
\medskip
\emph{$^{c}$LPTM, Universite de Cergy-Pontoise (CNRS UMR 8089), Saint-Martin2 \\
2 avenue Adolphe Chauvin, F-95302 Cergy-Pontoise Cedex, France}\\ 
\medskip
\emph{$^{p}$ Department of Physics, Brown University, \\
Box 1843, Providence, RI 02912, USA}\\
\medskip
E-mails: \emph{ianiceto@math.ist.utl.pt, Jean.Avan@u-cergy.fr \& Antal\_Jevicki@brown.edu}
\par\end{center}

%\begin{center}
%\bigskip  26 November 2009
%\par\end{center}
\begin{quote}
\bigskip \bigskip \textsf{\textbf{ Abstract:}}  We examine the Hamiltonian structures of some
Calogero-Moser and Ruijsenaars-Schneider
$N$-body integrable models. We propose explicit formulations of the bihamiltonian
structures for the discrete models, and field-theoretical realizations of these structures.
We discuss the relevance of these realizations as collective-field theory
for the discrete models.

 \bigskip  \thispagestyle{empty}
\end{quote}

\section{Introduction\label{sec:Introduction}}

Bihamiltonian structures for $N$-body dynamical systems can be seen
as a dual formulation of integrability, in the sense that they substitute
a hierarchy of compatible Poisson structures to a hierarchy of commuting
Hamiltonians, to establish Liouville integrability of a given system \cite{Magri:1978,bartocci:2009} .
Our specific interest for this formulation stems here from the conjecture
that, in the case of $N$-body Ruijsenaars-Schneider (RS) model\cite{Ruijsenaars:1986vq}, its
higher Hamiltonian structures may be the relevant framework to describe
the dynamics of some magnon-type solutions of string
theory\cite{Aniceto:2006rr,Aniceto:2008pc}.
Relevance of higher Poisson structures were demonstrated in the associated
Sine-Gordon theory in\cite{Mikhailov:2005sy}.

This leads us to a general questioning of the bihamiltonian structures
for related integrable discrete $N$-body systems, and continuous realizations
thereof. More specifically: The explicit realization of the bihamiltonian structure for the rational
Calogero-Moser (CM) model \cite{Magri:1978,Borisov:1999,bartocci:2009}%
\footnote{We would like to thank one of the referees for pointing out \cite{Borisov:1999}.} is the
basis for our construction,
leading us towards our current proposition
of a bihamiltonian structure for $A_n$ rational Ruijsenaars-Schneider
models, and trigonometric Calogero-Moser models. 

At this point we wish to make an important remark: a realization of a bihamiltonian structure was
proposed long ago for the Relativistic N-body Toda model (see e.g. \cite{Suris:1993,Damianou:1994})
which is a long-range limit of Ruijsenaars-Schneider dynamics. However the difficulty
in the full Calogero-Moser and Ruijsenaars-Schneider case lies (in technical
terms) in the dynamical nature of the $r$-matrix structure which precludes the
use of canonical definitions a la Sklyanin (such as discussed in \cite{Suris:1993})
of the second Hamiltonian structure as a direct ``quadratization''
of the first Hamiltonian structure encapsulated in any linear $r$-matrix structure
(see also \cite{Li:1989,Oevel:1989pa} and references therein).

In parallel
we propose a realization of these three bihamiltonian structures in terms
of continuous field theories, which can be identified, at least in
the two Calogero-Moser cases, with the collective-field continuous limit of the
discrete systems. Identification in the RS case is more questionable and shall be accordingly
dealt with in a further study; we shall only give some comments about it.

We shall successively describe the results for the rational CM, trigonometric
CM and rational RS models. We denote in the discrete case \textquotedbl{}bihamiltonian
structures\textquotedbl{} only those pairs of compatible Poisson brackets obeying in addition the 
hierarchy equation:\[
\left\{ h_{n},\mathcal{O}\right\} _{1}=\left\{ h_{n-1},\mathcal{O}\right\} _{2},\]
where $h_{n}$ are the tower of commuting Hamiltonians, and $\mathcal{O}$
is any observable.

The bihamiltonian structure for the discrete rational CM model is
described in \cite{Magri:1978} and further justified in \cite{bartocci:2009}
by explicit construction of the corresponding deformation of the canonical
$1$-form by a Nijenhuis-torsion free tensor. We give here an explicit realization
of the first two Poisson structures in terms of a collective field
$\alpha\left(x\right)$. The first one of these is the already known
collective-field formulation of the rational CM
\cite{Das:1990kaa,Avan:1991ik,Avan:1991kq}.
The formalism was recognized as being suitable for a useful representation
of higher conserved charges and symmetries of the N-body
system\cite{Avan:1991kq,Jevicki:1993qn}
For thee second Poisson bracket one  requires a deformation of the Poisson brackets
of $\alpha\left(x\right)$ together with a change in the realization of the
variables, understood from the change in the phase-space volume element
in the collective field formulation, precisely related to
the differing Poisson structure.

We then discuss the case of the trigonometric CM model. Based on the
identification between the second Poisson structures of the rational
CM and first Poisson structures of the trigonometric CM, we propose
a second Poisson structure for the trigonometric CM. A consistent
formulation in the framework of a continuous field theory is proposed
in terms of a collective field $\alpha\left(x\right)$ . The validity of 
the hierarchy equation for the corresponding two brackets is 
conjectured in the discrete case from consistency checks on the continuous
realization.

We finally address the case of rational RS model. The first Poisson
structure on discrete observables was derived recently \cite{Ayadi:2009};
we propose here a direct formulation from the Lax matrix Poisson
structure and its key $r,s$-matrix formulation.
Once again the identification of this Poisson structure with the second
Poisson structure of rational CM model allows us to propose a second
Poisson structure for the rational RS model, with the hierarchy
property. We then construct a field-theoretical realization of this
bihamiltonian structure. Its relevance as a collective
field theory for rational RS is, as we have indicated, a delicate issue,
essentially postponed until further studies.

All matrix indices throughout this paper are taken to vary between $1$
and $N$ for $N$ a given finite integer.

\section{Bihamiltonian Structure for Rational Calogero-Moser\label{sec:Bihamiltonian-Structure-rat-CM}}

This was derived in \cite{Magri:1978,bartocci:2009}. It is expressed
directly in terms of observables, respectively $I_{k}\equiv\frac{1}{k}\mbox{tr}\left(L^{k}\right)$
and $J_{\ell}=\mbox{tr}\left(L^{\ell-1}Q\right)$, where $L$ is the
Lax matrix and $Q$ is the position matrix :\begin{equation}
L_{ij}=p_{i}\delta_{ij}+\frac{g}{\left(q_{i}-q_{j}\right)}\left(1-\delta_{ij}\right),\; Q=\mbox{diag}\left(q_{i}\right).\label{eq:Lax-matrix-rational-CM}\end{equation}
From the first canonical Poisson bracket $\left\{ p_{i},q_{j}\right\}_1 =\delta_{ij}$
one gets the first Poisson bracket expression for the invariant variables
$I_{k},\, J_{\ell}$:\begin{flalign}
\left\{ I_{k},I_{m}\right\} _{1} & =0,\nonumber \\
\left\{ I_{k},J_{\ell}\right\} _{1} & =-\left(k+\ell-2\right)I_{k+\ell-2},\label{eq:First-PB-rational-CM}\\
\left\{ J_{k},J_{\ell}\right\} _{1} & =\left(\ell-k\right)J_{k+\ell-2}\,.\nonumber \end{flalign}
The second bracket is obtained directly by exploiting the reduction
scheme yielding $L$ and $Q$ from the original matrix variables,
and the construction of an explicit Nijenhuis-torsion free tensor yielding
the second Poisson bracket of $T^{*}\mathfrak{gl}\left(n\right)$.
It reads:\begin{flalign}
\left\{ I_{k},I_{m}\right\} _{2} & =0,\nonumber \\
\left\{ I_{k},J_{\ell}\right\} _{2} & =-\left(k+\ell-1\right)I_{k+\ell-1},\label{eq:Second-PB-rational-CM}\\
\left\{ J_{k},J_{\ell}\right\} _{2} & =\left(\ell-k\right)J_{k+\ell-1}.\nonumber \end{flalign}
It is not easy to express $\left\{ ,\right\} _{2}$ in terms of the
$p,q$ variables, although it may be a very useful alternative in
view of the extension to the trigonometric CM or rational RS models.

\paragraph{Remark }

It is easy to check (directly) that these two compatible Poisson bracket
structures are in fact one pair amongst any one chosen in the following
set:\begin{flalign}
\left\{ I_{k},I_{m}\right\} _{a} & =0,\nonumber \\
\left\{ I_{k},J_{\ell}\right\} _{a} & =-\left(k+\ell-2+a\right)\left(1+\frac{\lambda_{a}}{k}\right)I_{k+\ell-2+a},\label{eq:General-PB-rational_CM}\\
\left\{ J_{k},J_{\ell}\right\} _{a} & =\left(\ell-k\right)J_{k+\ell-2+a},\nonumber \end{flalign}
where $a$ is any integer in $\mathbb{Z}$ and $\lambda_{a}$ an arbitrary
$c$-number. Indeed one has:
\begin{description}
\item[{Theorem$\;1$}] Any linear combination $\left\{ ,\right\} _{a}+x\left\{ ,\right\} _{a'}$
with $a\ne a',\, x\in\mathbb{C}$, yields a skew-symmetric associative
Poisson bracket.
\end{description}
One has here a one-parameter ($\lambda_{a}$) multihamiltonian structure
when $a\in\mathbb{Z}$. More general mixed brackets $\left\{ I_{k},J_{\ell}\right\} _{a}$
may be derived but we have not solved the general coboundary equation
associated to it.

It will be important soon to specify the third Hamiltonian structure
of the hierarchy starting with $\left\{ ,\right\} _{0}$ and $\left\{ ,\right\} _{1}$.
It can be directly computed using the explicit recursion operator
in \cite{bartocci:2009}. It is unambiguously found to be given by
$\left\{ ,\right\} _{a}$ with $a=3$ and $\lambda_{3}=0$.

\section{Realization of the Bihamiltonian Structure: Collective Field Theory\label{sec:Bihamiltonian-Structure-Coll-FT}}

The collective field theory describing the $N\rightarrow\infty$ continuous
limit of the $N$-site CM model was described in \cite{Jevicki:1991yi}. 
It is obtained as the result of a phase-space integral, over
the continuous version of variables $p$ and $q$, replacing the discrete traces
of polynomials of the Lax matrix $L$ (substituted consistently by
$p\mbox{\ensuremath{\left(x\right)}}$) and position matrix $Q$
(substituted by $q\mbox{\ensuremath{\left(x\right)}}$). The dynamical
variables $\alpha^{\pm}$ are identified with the end-points of the
$p$-integration. Their Poisson bracket structure must be determined
by consistency with the original Poisson bracket structure of the
discrete traces, precisely $I_{k}$ and $J_{\ell}$. The phase space
integration however implies a subtle redefinition of the observables,
when higher Hamiltonian structures are to be represented, since the
invariant phase-space volume is accordingly redefined.

The first Poisson structure is described by \cite{Avan:1991kq}
:\begin{equation}
\begin{array}{cll}
I_{k} & =\int^{\alpha}dp\, dq\,\frac{p^{k}}{k} & \equiv\int dx\,\frac{\alpha^{k+1}}{k(k+1)}\,,\\
J_{\ell} & =\int^{\alpha}dp\, dq\, q\cdot p^{\ell-1} & \equiv\int dx\, x\frac{\alpha^{\ell}}{\ell}\,,\end{array}\label{eq:Obs-I-J-in-phase-space}\end{equation}
with the Poisson bracket structure for $\alpha$ given by the first
Poisson structure in KdV:\begin{equation}
\left\{ \alpha\left(x\right),\alpha\left(y\right)\right\} _{1}=-\delta'\left(x-y\right)\label{eq:KdV-First-PB}\end{equation}
It is immediate to check that it yields precisely the Poisson brackets
$\left\{ ,\right\} _{1}$.

To obtain the realization of the second Poisson structure in terms
of \textquotedbl{}collective\textquotedbl{} fields we assume that
the collective variables $I_{k}$ and $J_{\ell}$ are obtained by
a similar integration, over a modified phase space volume, taking
into account the change in Poisson brackets of the same densities
$I_{k},\, J_{\ell}$ in terms of $p$ and $q$. In particular we assume
that the degree in $p$ of the density yielding respectively $I_{k}$
and $J_{k}$ again differ by one unit. We are thus lead to the
following general form for the observables. \begin{flalign}
I_{k} & =\int dx\,\alpha\left(x\right)^{k+a}f\left(k\right),\nonumber \\
J_{\ell} & =\int dx\, x\,\alpha\left(x\right)^{\ell+a-1}g\left(\ell\right)\label{eq:I-and-J-in-terms-of-alpha-general}\end{flalign}
and Poisson structure for $\alpha$, assumed to be polynomial symmetric
in $\alpha$:\begin{equation}
\left\{ \alpha\left(x\right),\alpha\left(y\right)\right\} =-\alpha\left(x\right)^{c/2}\alpha\left(y\right)^{c/2}\delta'\left(x-y\right)\label{eq:PB-alphas-with-general-proportionality}\end{equation}
Determination of the numbers $a,c,f(k)$ and $g(\ell)$ follows straightforwardly
from plugging \eqref{eq:I-and-J-in-terms-of-alpha-general} and \eqref{eq:PB-alphas-with-general-proportionality}
into \eqref{eq:Second-PB-rational-CM}, yielding the following results
up to an overall normalization of all $k$-indexed observables by
a factor $\lambda^{k-1}$ with arbitrary $\lambda$ (corresponding to
an arbitrary renormalisation of $\alpha$).

The second Poisson structure $\left\{ ,\right\} _{2}$ is realized by:
\[
\begin{array}{cll}
I_{k} & =\int^{\alpha}p^{-1}dp\, dq & \equiv\int dx\,\frac{\alpha^{k}}{k^{2}}\,,\\
J_{\ell} & =\int^{\alpha}p^{-1}dp\, dq\, qp^{\ell-1} & \equiv\int dx\, x\,\frac{\alpha^{\ell-1}}{\ell-1}\,,\end{array}\]
with the following Poisson brackets for $\alpha$:\[
\left\{ \alpha\left(x\right),\alpha\left(y\right)\right\}_{2} =\alpha\left(x\right)\alpha\left(y\right)\delta'\left(x-y\right).\]

Notice that the result for the continuous observables is indeed obtained
by a change in the phase space volume $dp\, dq\rightarrow p^{-1}dp\, dq$.
Accordingly the canonical discrete variable becomes now $\ln p$,
and one consistently finds that it is now $\ln\alpha\left(x\right)$ which
(in the continuous limit) has a canonical Poisson bracket structure. This Poisson
bracket is the third in the KdV hierarchy. It thus seems that the second
Poisson bracket of KdV $\left\{ \alpha\left(x\right),\alpha\left(y\right)\right\} \sim\alpha(x)^{1/2}\alpha(y)^{1/2}\delta'(x-y)$
does not play a role in the CM framework.%
\footnote{Here we are referring to the long wavelength limit of the second Poisson bracket, not the full Poisson bracket. \label{foot:long-wavelength}}

Also notice that although the second discrete Poisson bracket
does realize the hierarchy property and is therefore correctly identified as
THE second Poisson bracket in the rational CM bihamiltonian hierarchy
the \textquotedbl{}second\textquotedbl{} continuous Poisson bracket
is not so, since not only the field bracket but also the definition
of the observables has to be changed. If for consistency one
compute the Poisson bracket of the same variables in terms of continuous
fields, it yields instead\[
\left\{ h_{n},\int dx\, x\,\frac{\alpha^{\ell-1}}{\ell-1}\right\} _{2}\equiv\left\{ h_{n+2},\int dx\, x\,\frac{\alpha^{\ell-1}}{\ell-1}\right\} _{1},\]
exhibiting a shift of $2$ in the degree of Hamiltonian,
from which one inescapably concludes that the continuous realization
of the second discrete Hamiltonian structure for rational CM is in fact a
third Hamiltonian structure for the collective field theory. The second
Hamiltonian structure of the latter corresponds obviously to the second KdV
bracket, and is seemingly (as we have said) not manifest in the discrete
CM frame.

\section{Trigonometric Calogero-Moser Model\label{sec:Trigonometric-Calogero-Moser-Model}}

An algebra of observables for the discrete CM trigonometric model
is written \cite{Avan:1994us} in terms of the coordinate matrix%
\footnote{We should introduce some notation here. The matrix $e_{ij}$ is a
matrix which has all elements zero except the element $ij$.%
} $K=\sum_{j}\exp(q_{j})\, e_{jj}$ and Lax matrix $L=\sum_{i}p_{i}e_{ii}+\sum_{i\ne j}g\frac{\cos(q_{i}-q_{j})}{\sin(q_{i}-q_{j})}e_{ij}$
using the first canonical Poisson structure $\left\{ p_{i},q_{j}\right\} =\delta_{ij}$.
The overcomplete set of observables \[
\left\{ W_{mn}=\mbox{Tr}L^{m}e^{nQ},\, m\ge0,\, n\ge0\right\} \]
is easily shown to realize a $W_{1+\infty}$ algebra (albeit in a very
degenerate representation due to the existence of
algebraic relations between the $W_{mn}$ issuing from their realization as $NxN$ matrices):\[
\left\{ W_{mn},W_{pq}\right\} _{1}=\left(mq-np\right)W_{m+p-1,n+q}+\mbox{lower-order terms.}\]

In order to define a second Poisson structure, following the previous
derivation, we shall use as independent
variables not the $\left(p_{i},q_{j}\right)$ but a subset of algebraically
independent observables from the set $\left\{ W_{mn}\right\} $ such
that the change of variables be bijective (at least from a given Weyl
chamber for the position and momenta variables, since the discrete permutation over
indices is factored out by the use of invariant traces). Guided by the discussion
in \cite{bartocci:2009} we see that either $\left\{ W_{m0},W_{m1},\, m\le N\right\} $
or $\left\{ W_{0m},W_{1m},\, m\le N\right\} $ provide such a subset. Using the
first subset seems a priori natural since it contains the Hamiltonians $W_{l,0}$ .
However $\left\{ W_{m1},W_{p1}\right\} _{1}=\left(mq-np\right)W_{m+p-1,2}+\mbox{lower-order terms.}$.
It is in principle possible to re-express $W_{m+p-1,2}$ in terms of $W_{k,1}$ and $W_{l,0}$
since these second-index $0$ and $1$ observables provide an algebraically complete set of new variables. However
this re-expression is expected to be quite cumbersome: in particular it will 
certainly yield non-linear expressions, suggesting that a consistent guess of compatible second Poisson
bracket will be difficult to formulate.

The second set however closes linearly and explicitly under the first Poisson bracket
and it is thus this one which we choose to define the Poisson hierarchy.
It is also crucial to note that no lower-order term appears in its
Poisson brackets. It then turns out by simple inspection that the
first Poisson structure for trigonometric CM expressed in terms of
variables $W_{0,1;m}$ is isomorphic to the second Poisson structure
for rational CM. It thus seems natural to propose as a second Poisson
structure for trigonometric CM the third Poisson structure of rational
CM. In terms of $W$-variables it easily reads:\begin{equation}
\left\{ W_{im},W_{jn}\right\} _{2}=\left(in-jm\right)W_{i+j-1,m+n+1}\,.\label{eq:Second-PB-trig-CM}\end{equation}

This characterizes $\left\{ ,\right\} _{1}$ and $\left\{ ,\right\} _{2}$
as a pair of compatible Poisson structures for trigonometric CM model.
However, in order to further characterize $\left\{ ,\right\} _{1}$ and $\left\{ ,\right\} _{2}$
as a bihamiltonian structure for the trigonometric CM model, we need
to prove that it realizes the hierarchy equality for
evolution of observables:\[
\left\{ W_{m0},W_{in}\right\} _{2}\equiv\left\{ W_{m+1,0},W_{in}\right\} _{1}\,,\: i=0,1,\, m\le N\,.\]
This is not easy since it implies that one is able to compute the
second Poisson bracket ot the variables $W_{m0}$, once again a difficult task
given that they are redundant variables and we do not control the lower-order
terms. We shall now use the collective
field description of the continuous limit to at least establish the
consistency of this statement.

\section{Realization: Continuous Trigonometric Calogero-Moser Model\label{sec:Continuous-Trigonometric-Calogero-Moser}}

It is known that for a particular value of the coupling constant the
trigonometric CM model is equivalent, at the continuum level, to a
free fermion on a circle \cite{Jevicki:1991yi}. This suggests that the collective
field theory for trigonometric CM should again be expressed as a phase
space integral, this time over a circle in the $q$ variable, yielding the realization
of the first Poisson structure as:\[
\begin{array}{lll}
W_{0m}=\mbox{Tr}e^{mQ} & \mbox{becomes} & W_{0m}=\int dx\, e^{mx}\alpha\left(x\right),\\
W_{1m}=\mbox{Tr}e^{mQ}L & \mbox{becomes} & W_{1m}=\int dx\, e^{mx}\frac{\alpha\left(x\right)^{2}}{2}\,,\end{array}\]
and generically:\[
W_{nm}=\mbox{Tr}e^{mQ}L^{n}\quad\mbox{becomes}\quad W_{nm}=\int dx\, e^{mx}\frac{\alpha\left(x\right)^{n+1}}{n+1}\]
with the Poisson bracket $\left\{ \alpha\left(x\right),\alpha\left(y\right)\right\} _{1}=\delta'\left(x-y\right)$.

This set of integrated collective-field densities realizes indeed
the leading (linear) order of the Poisson bracket algebra for the
discrete $W_{mn}$ generators under the first Poisson bracket. Note
that a similar property already held in the rational case, when one
extended the Poisson algebra to the redundant discrete generators
$\mbox{Tr}L^{m}Q^{n}$, realized in the continuum limit as $\int dx\, x^{m}\frac{\alpha^{n+1}}{n+1}.$

Realization of the second Poisson structure is, strictly speaking,
only available at this stage for the generators .$W_{0m},\, W_{1m}$
We assume as a generic form for this realization the following monomial
integrals:\[
W_{im}=\int dx\, e^{(m+a)x}\frac{\alpha\left(x\right)^{i+1+b}}{i+1+b}.\]
Indeed this is the only way to guarantee that the separate additivity (up to
a constant!) of the indices $i$ and $m$ will be preserved
in the formulation of the Poisson algebra. The Poisson structure for
the field $\alpha$ is taken to be the most generic symmetric monomial
expression in $\alpha$ and $e^x$ \[
\left\{ \alpha\left(x\right),\alpha\left(y\right)\right\} =e^{\frac{c}{2}(x+y)}\left(\alpha(x)\alpha(y)\right)^{d/2}\delta'\left(x-y\right).\]

Plugging these ans\"{a}tze for $W_{0m},\, W_{1m}$ into the expected
algebraic structure yields a unique answer:\[
a=-1,\, c=2,\, b=0,\, d=0.\]
In particular one remarks that it is the new $\tilde{\alpha}\left(x\right)\equiv e^{-x}\alpha\left(x\right)$
which now realizes a canonical Poisson bracket $\left\{ \tilde{\alpha}\left(x\right),\tilde{\alpha}\left(y\right)\right\} _{2}=\delta'\left(x-y\right)$.

Because this realization is unique, and completely determined by the
Poisson brackets of the independent generators $W_{0m},W_{1m}$, it
seems acceptable to conjecture that it will entail a similar realization
for the redundant higher-order generators $W_{nm},\, n\ge2$.
From our previous conjecture they are represented as:\[
W_{nm}=\int dx\, e^{(m-1)x}\frac{\alpha\left(x\right)^{n+1}}{n+1}\,.\]

We can now compute at least the leading order of the actual Hamiltonian
action on these conjectured continuous observables, implied by the
second Poisson structure:\[
\left\{ W_{n0},W_{im}\right\} ^{\mathrm{continuous}}_{(2)}=nm\, W_{n+i-1,m+n+1}.\]
If as we have conjectured, this representation is indeed the continuous
representation of the second Poisson structure on all the observables
of the trigonometric CM model, this equation guarantees that, at the
discrete level, we have:\[
\left\{ W_{n0},W_{im}\right\} ^{\mathrm{discrete}}_{(2)}=nm\, W_{n+i-1,m+n+1}=\left\{ W_{n+1,0},W_{im}\right\} ^{\mathrm{discrete}}_{(1)}\]
up to lower-order terms, which are in any case not accessible to the
continuous representation. Therefore is is not inconsistent to characterize
{\small $\left\{ ,\right\} ^{\mathrm{discrete}}_{(2)}$ }as a second
Hamiltonian structure in a multihamiltonian hierarchy for the trigonometric
CM.

\section{Bihamiltonian Structure for Rational Ruijsenaars-Schneider Model\label{sec:Bihamiltonian-Structure-rat-RS}}

A consistent construction of a bihamiltonian structure can be
formulated on the following lines.
%\begin{description}
%\item [{a.}] 
\vspace{10pt}

\hspace{3pt} \textbf{a.} $\,\,$ The canonical Poisson structure in terms of the basic variables
$p$ and $q$ is again re-expressed as a Poisson structure for the following variables:\begin{equation}
I_{k}=\mbox{Tr}\frac{L^{k}}{k}\,,\; J_{\ell}=\mbox{Tr}\, QL^{\ell-1},\end{equation}
where $L$ is the Lax matrix for rational RS and $Q=\mathrm{diag}\,(q_{i})$
as before. 
Direct derivation of the Poisson structure for these observables now follows from
the $r$ matrix structure of the rational RS Lax matrix $L$.
It is given
by \begin{equation}
L=\sum_{k,j=1}^{N}\frac{\gamma}{q_{k}-q_{j}+\gamma}b_{j}\,\e_{kj}\,,\quad b_{k}=e^{p_{k}}\prod_{j\ne k}\left(1-\frac{\gamma^{2}}{(q_{k}-q_{j})^{2}}\right)^{1/2}.\label{eq:Lax-matrix-for-rational-RS}\end{equation}

The matrix $\mathbf{e}_{kj}$ is the $N\times N$ matrix
with all components zero except the $kj$ component, which is one.

The canonical Poisson bracket in the canonical variables
$q_{k},p_{j}$: \begin{equation}
\left\{ p_{k},p_{j}\right\} _{0}=\left\{ q_{k},q_{j}\right\} _{0}=0\,,\quad\left\{ q_{j},p_{k}\right\} _{0}=\delta_{kj}\,.\label{eq:Canonical-PB-in-p-q}\end{equation}
becomes, in the $q_{k},b_{j}$ variables:\begin{align}
\left\{ q_{k},q_{j}\right\}  & =0,\nonumber \\
\left\{ q_{k},b_{j}\right\}  & =b_{k}\delta_{kj}\,,\label{eq:Canonical-PB-in terms-of-qs-and-bs}\\
\left\{ b_{k},b_{j}\right\}  & =\left\{ \frac{1}{q_{j}-q_{k}+\gamma}-\frac{1}{q_{k}-q_{j}+\gamma}+\frac{2(1-\delta_{kj})}{q_{k}-q_{j}}\right\} b_{k}b_{j}.\nonumber \end{align}

This Poisson bracket is quadratic in the Lax matrix \cite{Suris:1996mg}
$L$: \begin{equation}
\left\{ L_{1}\overset{\ot}{,}L_{2}\right\} =a_{12}L_{1}L_{2}-L_{1}L_{2}d_{12}-L_{1}s_{12}L_{2}+L_{2}s_{21}L_{1}\,.\label{eq:First-PB-of two-Lax-matrices}\end{equation}
where:\begin{align}
d_{12} & =-a_{12}^{CM}-w,\nonumber \\
a_{12} & =-a_{12}^{CM}-s_{12}^{CM}+s_{21}^{CM}+w,\label{eq:Definition-of-the-RS-tensors}\\
s_{21} & =s_{12}^{CM}-w,\nonumber \\
s_{12} & =s_{21}^{CM}+w,\nonumber \end{align}
The following tensors already existed in the Calogero-Moser
case \cite{Avan:1992km}:%
\footnote{For the rational CM model we have the Lax matrix 
\begin{equation}
L^{r}=\sum_{k=1}^{N}p_{k}\,\mathbf{e}_{kk}+\sum_{k\ne j}^{N}\frac{\gamma}{q_{k}-q_{j}}\,\mathbf{e}_{kj}\,,\label{eq:Lax-matrix-rational-CM}\end{equation}
}\begin{align}
a_{12}^{CM} & =-\sum_{k\ne j}\frac{1}{q_{j}-q_{k}}\,\mathbf{e}_{jk}\otimes\mathbf{e}_{kj}\,,\label{eq:a-s-tensors-for-rat-CM}\\
s_{12}^{CM} & =\sum_{k\ne j}\frac{1}{q_{j}-q_{k}}\,\mathbf{e}_{jk}\otimes\mathbf{e}_{kk}\,.\nonumber \end{align}

They actually also define\cite{Suris:1996mg} the famous non-skew symmetric dynamical $r$-matrix of the rational
CM model by \begin{equation}
r_{12}^{CM}=a_{12}^{CM}+s_{12}^{CM}\,.\label{eq:r-matrix-for-CM}\end{equation}

The tensor $w$ in \eqref{eq:Definition-of-the-RS-tensors} only appears
in the RS model, it is defined by \[
w=\sum_{k\ne j}\frac{1}{q_{k}-q_{j}}\e_{kk}\ot\e_{jj}.\]
Finally one can see that the tensors in \eqref{eq:Definition-of-the-RS-tensors}
obey the classical consistency relation \begin{equation}
a_{12}-d_{12}+s_{21}-s_{12}=0.\label{eq:Relation-between-RS-tensors}\end{equation}
which we shall see to be necessary for Poisson-commutation of the traces.

We also determine the Poisson brackets of the Lax operator with
the position operator $Q$, defined as\begin{equation}
Q=\sum_{k}q_{k}\e_{kk}\,,\label{eq:Q-operator-for-rat-CM}\end{equation}
obtaining\begin{align}
\left\{ L_{1}\overset{\ot}{,}Q_{2}\right\}  & =\sum_{i,j,k}\left\{ L_{ij},q_{k}\right\} \e_{ij}\ot\e_{kk}=\sum_{i,j,k}\frac{\gamma}{q_{i}-q_{j}+\gamma}\left\{ b_{j},q_{k}\right\} \e_{ij}\ot\e_{kk}\nonumber \\
 & =-\sum_{i,j}L_{ij}\e_{ij}\ot\e_{jj}=-L_{1}^{r}\cdot\sum_{j}\e_{jj}\ot\e_{jj}\,.\label{eq:PB-of-Lax-matrix-and-Q-for-rat-RS}\end{align}

We now re-write the above Poisson brackets using
as basic variables the following traces: \begin{equation}
W_{n}^{m}=\tr\left(L^{n}Q^{m}\right),\; m=0,1\,.\label{eq:Definition-of-traces-of-LnQm}\end{equation}
The simplest of the Poisson brackets is:\begin{align*}
\left\{ W_{n}^{0},W_{m}^{0}\right\}  & =\tr_{1,2}\sum_{i,j=1}^{n,m}\left\{ L_{1}\overset{\ot}{,}L_{2}\right\} L_{1}^{n-1}L_{2}^{m-1}\\
 & =m\, n\,\tr_{1,2}\left((a_{12}-d_{12}-s_{12}+s_{21})L_{1}^{n}L_{2}^{m}\right)=0,\end{align*}
where we used the key consistency relation \eqref{eq:Relation-between-RS-tensors}.

The next Poisson bracket to be determined is \[
\left\{ W_{n}^{0},W_{m}^{1}\right\} =\underbrace{\tr_{1,2}\left(\sum_{i,j}\left\{ L_{1}\overset{\ot}{,}L_{2}\right\} L_{1}^{n-1}L_{2}^{j-1}Q_{2}L_{2}^{m-j}\right)}_{A_{01}}+\underbrace{\tr_{1,2}\left(\sum_{i=1}^{n}L_{2}^{m}L_{1}^{n-i}\left\{ L_{1}\overset{\ot}{,}Q_{2}\right\} L_{1}^{i-1}\right)}_{B_{01}}.\]
The first term is re-written as \begin{align*}
A_{01} & =\tr_{1,2}\left\{ n\left(a_{12}-s_{12}\right)L_{1}^{n}\left[L_{2}^{m},Q_{2}\right]+\sum_{j=1}^{m}L_{2}^{j}Q_{2}L_{2}^{m-j}\left[s_{12},L_{1}^{n}\right]\right\} .\end{align*}
We once again made use of the cyclicity of the trace and of the relation
\eqref{eq:Relation-between-RS-tensors}. If we now use the explicit
formulas \eqref{eq:Definition-of-the-RS-tensors}, we find that \[
a_{12}-s_{12}=-a_{12}^{CM}-s_{12}^{CM}=-r_{12}^{CM},\]
where the superscript $CM$ corresponds to the Calogero-Moser model.
Then we write\begin{align*}
\tr_{1,2}\left(\sum_{j=1}^{m}L_{2}^{j}Q_{2}L_{2}^{m-j}\left[s_{12},L_{1}^{n}\right]\right) & =\sum_{j=1}^{m}\left(L^{j}QL^{m-j}\right)_{lk}\left(s_{12}\right)_{i'j'kl}L_{mn}^{n}\left(\delta_{j'm}\delta_{i'n}-\delta_{i'n}\delta_{j'm}\right)=0.\end{align*}
This allows us to simplify $A_{01}$ even further:\begin{align*}
A_{01} & =\tr_{1,2}\left(n(-r_{12}^{CM})L_{1}^{n}\left[L_{2}^{m},Q_{2}\right]\right)=-n\left(L_{ji}^{n}(r_{12}^{CM})_{ijml}L_{lm}^{m}\left(Q_{mm}-Q_{ll}\right)\right).\end{align*}
Using the expression for $r_{12}^{CM}$ in components
we finally find \[
A_{01}=-n\sum_{m\ne l}L_{ml}^{n}L_{lm}^{m}=-n\tr\left(L^{m+n}\right)+n\sum_{k}L_{kk}^{n}L_{kk}^{m}\,.\]
Let us now turn to the second term of the Poisson brackets $B_{01}$:\begin{align*}
B_{01} & =\tr_{1,2}\left(n\, L_{2}^{m}\left\{ L_{1}\overset{\ot}{,}Q_{2}\right\} L_{1}^{n-1}\right)=-n\sum_{j}L_{jj}^{n}L_{jj}^{m}\,.\end{align*}
The final result for this Poisson bracket is just \begin{equation}
\left\{ W_{n}^{0},W_{m}^{1}\right\} =-n\tr\left(L^{m+n}\right)=-nW_{m+n}^{0}\,.\end{equation}

The final Poisson bracket to determine is \begin{align*}
\left\{ W_{n}^{1},W_{m}^{1}\right\}  & =\underbrace{\tr_{1,2}\left(\sum_{i,j=1}^{n,m}\left\{ L_{1}\overset{\ot}{,}L_{2}\right\} L_{1}^{i-1}Q_{1}L_{1}^{n-i}L_{2}^{j-1}Q_{2}L_{2}^{m-j}\right)}_{A_{11}}+\\
 & +\underbrace{\tr_{1,2}\left(\sum_{j=1}^{m}L_{1}^{n}\left\{ Q_{1}\overset{\ot}{,}L_{2}\right\} L_{2}^{j-1}Q_{2}L_{2}^{m-j}\right)}_{B_{11}}+\underbrace{\tr_{1,2}\left(\sum_{i=1}^{n}L_{2}^{m}\left\{ L_{1}\overset{\ot}{,}Q_{2}\right\} L_{1}^{i-1}Q_{1}L_{1}^{n-i}\right)}_{C_{11}}.\end{align*}
First of all $A_{11}$ is\begin{align*}
A_{11} & =\tr_{1,2}\left(\sum_{i=1}^{n}\left(a_{12}-s_{12}\right)L_{1}^{i}Q_{1}L_{1}^{n-i}\left[L_{2}^{m},Q_{2}\right]+\sum_{j=1}^{m}\left(a_{12}+s_{21}\right)\left[L_{1}^{n},Q_{1}\right]L_{2}^{j}Q_{2}L_{2}^{m-j}\right)+\\
 & \quad+\tr_{1,2}\left((\left[Q_{1},d_{12}\right]Q_{2}+\left[Q_{2},d_{12}\right]Q_{1}+Q_{2}\left[Q_{1},s_{12}\right]-Q_{1}\left[Q_{2},s_{21}\right])L_{1}^{n}L_{2}^{m}\right).\end{align*}
To simplify this expression, we need a few extra results. The first
one is\[
\tr_{1,2}\left((\left[Q_{1},d_{12}\right]Q_{2}+\left[Q_{2},d_{12}\right]Q_{1}\right)=0\]
due to the cyclicity of the trace. The second one is \begin{align*}
\tr_{1,2}\left(Q_{2}\left[Q_{1},s_{12}\right]L_{1}^{n}L_{2}^{m}\right) & =\left(L^{m}Q\right)_{ij}\left(s_{12}\right)_{klji}L_{lk}^{n}\left(Q_{kk}-Q_{ll}\right)=0,\end{align*}
where we have used that $s_{12}=s_{21}^{CM}+w$ from \eqref{eq:Definition-of-the-RS-tensors}:
the $w$ contribution is zero, because this tensor is diagonal on
both spaces 1 and 2; the contribution from $s_{21}^{CM}$ is also
zero due to this tensor being diagonal in the first space $1$. A
very similar result can be obtained for:\[
\tr_{1,2}\left(Q_{1}\left[Q_{2},s_{21}\right])L_{1}^{n}L_{2}^{m}\right)=0,\]
but in this case one would need to use $s_{21}=s_{12}^{CM}-w$ from
\eqref{eq:Definition-of-the-RS-tensors}.

With these results, $A_{11}$ boils down to \[
A_{11}=\tr_{1,2}\left(-\sum_{i=1}^{n}r_{12}^{CM}L_{1}^{i}Q_{1}L_{1}^{n-i}\left[L_{2}^{m},Q_{2}\right]+\sum_{j=1}^{m}r_{21}^{CM}\left[L_{1}^{n},Q_{1}\right]L_{2}^{j}Q_{2}L_{2}^{m-j}\right).\]
In this last expression we again used the relations directly derived
from \eqref{eq:Definition-of-the-RS-tensors}\[
a_{12}-s_{12}=-r_{12}^{CM}\,,\quad a_{12}+s_{21}=r_{21}^{CM}.\]
The two terms in $A_{11}$ are further simplified by the use of
$r_{12}^{CM}$ in components:\begin{align*}
-\sum_{i=1}^{n}\tr_{1,2}\left(r_{12}^{CM}L_{1}^{i}Q_{1}L_{1}^{n-i}\left[L_{2}^{m},Q_{2}\right]\right) & =-\sum_{i=1}^{n}\sum_{k\ne l}\left(L^{i}QL^{n-i}\right)_{lk}L_{kl}^{m}\\
 & =-n\,\tr\left(QL^{m+n}\right)+\sum_{i=1}^{n}\sum_{k}\left(L^{i}QL^{n-i}\right)_{kk}L_{kk}^{m}\,,\end{align*}
and likewise\begin{align*}
\sum_{j=1}^{m}\tr_{1,2}\left(r_{21}^{CM}\left[L_{1}^{n},Q_{1}\right]L_{2}^{j}Q_{2}L_{2}^{m-j}\right) & =m\,\tr\left(QL^{m+n}\right)-\sum_{j=1}^{m}\sum_{k}\left(L^{j}QL^{m-j}\right)_{kk}L_{kk}^{n}\,.\end{align*}
Finally $A_{11}$ becomes simply\[
A_{11}=\left(m-n\right)\,\tr\left(QL^{m+n}\right)+\sum_{i=1}^{n}\sum_{k}\left(L^{i}QL^{n-i}\right)_{kk}L_{kk}^{m}-\sum_{j=1}^{m}\sum_{k}\left(L^{j}QL^{m-j}\right)_{kk}L_{kk}^{n}\,.\]

We still have to determine the other terms $B_{11}$ and $C_{11}$.
Let us proceed with $B_{11}$:\begin{align*}
B_{11} & =\sum_{j=1}^{m}\tr_{1,2}\left(L_{1}^{n}L_{2}\cdot\sum_{k}\e_{kk}\ot\e_{kk}\cdot L_{2}^{j-1}Q_{2}L_{2}^{m-j}\right)\\
 & =\sum_{j=1}^{m}\sum_{k}L_{kk}^{n}\left(L^{j}QL^{m-j}\right)_{kk}.\end{align*}
In order to obtain the last line, we have used the fact that \[
\sum_{k}L_{kk}^{n}\left(QL^{m}\right)_{kk}-\sum_{k}L_{kk}^{n}\left(L^{m}Q\right)_{kk}=\sum_{k}L_{kk}^{n}Q_{kk}L_{kk}^{m}-\sum_{k}L_{kk}^{n}L_{kk}^{m}Q_{kk}=0.\]
Turning to $C_{11}$ one similarly obtains\begin{align*}
C_{11} & =-\sum_{i=1}^{n}\sum_{k}L_{kk}^{m}\left(L^{i}QL^{n-i}\right)_{kk}.\end{align*}

We finally write the result for the Poisson bracket:\[
\left\{ W_{n}^{1},W_{m}^{1}\right\} =A_{11}+B_{11}+C_{11}=\left(m-n\right)\,\tr\left(QL^{m+n}\right)=\left(m-n\right)W_{m+n}^{1}\,.\]
Summarizing the results obtained for the Poisson brackets of the traces,
we have for the rational RS model:\begin{eqnarray}
\left\{ W_{n}^{0},W_{m}^{0}\right\} _{1} & = & 0\,,\nonumber \\
\left\{ W_{n}^{0},W_{m}^{1}\right\} _{1} & = & -nW_{m+n}^{0}\,,\label{eq:Full-algebra-rational-RS}\\
\left\{ W_{n}^{1},W_{m}^{1}\right\} _{1} & = & \left(m-n\right)W_{m+n}^{1}\,.\nonumber \end{eqnarray}
Renormalizing the variables $W_{n}^{0,1}$ to our variables
$I_{k},J_{\ell}$, by\[
I_{k}=\frac{1}{k}W_{k}^{0}\:,\quad J_{\ell}=W_{\ell-1}^{1}.\]
we get:\begin{flalign}
\left\{ I_{k},I_{\ell}\right\} _{1} & =0\,,\nonumber \\
\left\{ J_{\ell},I_{k}\right\} _{1} & =\left(k+\ell-1\right)I_{k+\ell-1}\,,\\
\left\{ J_{\ell},J_{m}\right\} _{1} & =(m-\ell)\, J_{m+\ell-1}.\nonumber \end{flalign}

Another derivation of these Poisson structure was recently given \cite{Ayadi:2009}, using the realization
of the RS model by KKS reduction\cite{Feher:2009dh}, thereby bypassing the explicit
use of $r$ matrix structure.

The key remark here is that this canonical (first) bracket for the
rational RS is isomorphic to the second bracket $\left\{ ,\right\} _{2}$
(with $\lambda_{2}=0$) for the rational CM. This is consistent with
the remark in \cite{bartocci:2009} on the formal equality of the canonical
symplectic form on $T^{*}GL\left(n,\mathbb{C}\right)$ yielding the first Poisson
structure of trigonometric CM model, with the relevant symplectic form yielding
the second bracket for the rational CM model; together with the well-known
Ruijsenaars duality between trigonometric CM and rational RS, certainly valid
at least when the first Poisson structures are considered in both
formulations.
%\item [{b.}] 
\vspace{10pt}

\hspace{3pt}\textbf{b.} $\,\,$ Even though a direct computation of the new
symplectic form deformed by a Nijenhuis-torsion free tensor (i.e. the new canonical $1$-form)
is not available for rational RS (lacking an obvious choice of such Nijenhuis-torsion free
tensor), we however prove, in view of the explicit computations
of Section 2, that the natural
second Poisson brackets for the rational RS hierarchy are expressed
in terms of the observables $I_{k},\, J_{\ell}$, by the form of the
third Poisson brackets for rational CM written there.\\
Precisely:\begin{flalign}
\left\{ I_{k},J_{\ell}\right\} _{2} & =0,\nonumber \\
\left\{ J_{\ell},I_{k}\right\} _{2} & =\left(k+\ell\right)I_{k+\ell},\\
\left\{ J_{\ell},J_{m}\right\} _{2} & =\left(m-\ell\right)J_{m+\ell+1}.\nonumber \end{flalign}

\begin{description}
\item [{Proof:}]~\end{description}
\begin{enumerate}
\item $\left\{ ,\right\} _{2}$ is compatible with $\left\{ ,\right\} _{1}$
as a Poisson bracket structure for the observables $I_{k},\, J_{\ell}$
of RS since the Jacobi identity equations for {\small $\left\{ ,\right\} _{2}+x\left\{ ,\right\} _{1}$}
are the same as for {\small $\left\{ ,\right\} _{3}^{\mathrm{CM}}+x\left\{ ,\right\} _{2}^{\mathrm{CM}}$}.
\item We have the following relation\[
\left\{ J_{k},I_{\ell}\right\} _{2}\equiv\frac{d^{(2)}}{dt_{\ell}}J_{k}=\left\{ J_{k},I_{\ell+1}\right\} _{1}=\frac{d^{(1)}}{dt^{\ell+1}}J_{k}\]
which now characterizes $\left\{ ,\right\} _{1}$, $\left\{ ,\right\} _{2}$  as a bona fide bihamiltonian
structure for the RS hierarchy, defined by the set of Hamiltonians $\left\{ I_{\ell}\right\} $.
\end{enumerate}
%\end{description}

\section{Field-theoretical Realization of the Ruijsenaars-Schneider Structures\label{sec:Field-theoretical-RS-structures}}

We now propose from first principles a field-theoretical realization
of the two (bihamiltonian) Poisson structures previously computed
for the rational RS models. The first bracket is realized as:\[
I_{k}=\int dq\,\frac{e^{k\alpha}}{k^{2}}\,,\quad J_{\ell}=\int dq\, q\,\frac{e^{(\ell-1)\alpha}}{\ell-1}\,,\]
with Poisson bracket $\left\{ \alpha\left(x\right),\alpha\left(y\right)\right\} =\delta'\left(x-y\right).$
The exponential representation in $\alpha$ is motivated by the existence
of the Ruijsenaars duality between rational RS and trigonometric CM
\cite{Ruijsenaars:1986pp} under exchange of the variables $p$, $q$. 
 Accordingly, it appears consistent
to assume a dualized ($x\leftrightarrow\alpha$) representation in the
continuum case for the Poisson structure.

The second Poisson structure is now realized in the continuum, following
a similar scheme as in the rational and trigonometric CM case. Assuming
that a representation purely in $e^{\alpha(x)}$ will hold for the
$p$ variables, one introduces
as an ansatz for the observables the generic form:\[
I_{k}=\int dq\,\frac{e^{(k+a)\alpha}}{k+a}\,,\; J_{\ell}=\int dq\, q\,\frac{e^{(\ell+a-1)\alpha}}{\ell+a-1}\]
and similarly for the Poisson bracket\[
\left\{ \alpha\left(x\right),\alpha\left(y\right)\right\} =e^{\frac{c}{2}\left(\alpha(x)+\alpha(y)\right)}\delta'\left(x-y\right).\]
The exponential form for $\alpha$ in the Poisson brackets is required
by the exponential form in $I_{k}$ and $J_{\ell}$, which must be
preserved under Poisson bracket, to yield again $I$- and $J$- generators.
Plugging these ans\"{a}tze in the second Poisson bracket structure unambiguously
yields $a=-1$ and $c=2$. 

From $c=2$ it is now seen that $\phi\left(x\right)\equiv e^{-\alpha\left(x\right)}$
is a canonical field, $\left\{ \phi\left(x\right),\phi\left(y\right)\right\} =\delta'\left(x-y\right)$.
As in the case of rational CM model, this field-theoretical realization
is better interpreted as a third Poisson bracket for the continuous
theory since one obtains again a Hamiltonian evolution with a shift
of $2$ units in the degree:\[
\left\{ h_{n},\mathcal{O}\right\} _{\mathrm{continuous}}^{(2)}\equiv\left\{ h_{n-2},\mathcal{O}\right\} _{\mathrm{continuous}}^{(1)},\]
setting $h_{n}\equiv\int dq\,\frac{e^{n\alpha}}{n}$ in both cases,
as consistency requires.

The issue is now whether this field-theoretical realization 
can be obtained directly as a genuine collective field theory for RS model.
This requires a re-writing of
the operators $I_{k},J_{\ell}$ from a collective field theory
perspective, and from that a determination of the Poisson structures that arise,
thus confirming our ans\"{a}tze for these. A collective expression for
$I_{1}$ is known \cite{Awata:1994xd}, in terms of quantum MacDonald
operators, and one would like to extend the analysis done in \cite{Awata:1994xd}
to higher conserved quantities $I_{k}$ and $J_{\ell}$, from their
expressions in components found in \cite{Ruijsenaars:1986pp}. Such
a generalization, however, is not trivial to obtain, because higher
powers of the Lax matrix make such calculations very cumbersome.

Before concluding let us remark that the case of trigonometric RS model is much
more problematic to deal with at this time, due to the difficulty of
defining a Poisson-closed complete subalgebra of observables which
could be used as suitable coordinates. In this case the first
Poisson structure of neither the
set $\left\{ W_{m0},W_{m1},\, m\le N\right\} $ nor the dual set 
$\left\{ W_{0m},W_{1m},\, m\le N\right\} $
closes linearly. Indeed one has:
$\left\{ W_{n}^{1},W_{m}^{1}\right\} _{1}  = \left(m-n\right)W_{m+n}^{2} + ....$ and 
$\left\{ W_{1}^{n},W_{1}^{m}\right\} _{1}  = \left(m-n\right)W_{2}^{n+m} + ....$
The difficulty which lead us to eliminate the choice of the set $\left\{ W_{m0},W_{m1},\, m\le N\right\} $
in the trigonometric CM case exists now for both sets.

\section{Summary}

To conclude, we summarize the results obtained here and the remaining
issues regarding the construction of multihamiltonian structures for
the $N$-body models, their realization in continuous field theories
and interpretation of those as collective field theories.
\begin{description}
\item [{$A_{n}$~Calogero-Moser,~rational:}] Bihamiltonian structure were
already known in discrete case \cite{Magri:1978,bartocci:2009}. A collective-field
realization is proposed, with consistent ``bihamiltonian''
structure and consistently modified phase space.
\item [{$A_{n}$~Calogero-Moser,~trigonometric:}] Multiple Poisson structures
have been established. Consistent collective-field realizations are proposed, with
consistent ``bihamiltonian'' structure.
The hierarchy equations for the multiple discrete Poisson structures have not
been rigorously established but pass consistency checks.
\item [{$A_{n}$~Ruijsenaars-Schneider,~rational:}] A bihamiltonian structure
is established in the discrete case. A continuum realization is proposed, with
bihamiltonian structure; identification
as collective field theory is yet unproven.\\

\end{description}

\section*{Acknowledgements}

This work was partially funded by CNRS (J.A.) and the Department of Energy under contract DE-FG02-91ER40688 (I.A. and A.J.); I.A. was also partially supported by the Funda\c{c}\~{a}o para a Ci\^{e}ncia e a Tecnologia (FCT / Portugal). J.A. wishes to warmly thanks Brown University  Physics Department for their hospitality. I.A. would like to thank A. Jevicki for the support for this work during the summer of 2009.

The authors would also like to thank one of the referees and V. Roubtsov for their clarification of the nature of the second Poisson bracket of the Calogero-Moser model, found in footnote \ref{foot:long-wavelength}.

\bibliographystyle{utcaps_changed}
\bibliography{My_papers}

\end{document}